\documentclass[aps,prd,reprint,groupedaddress,nofootinbib,floatfix]{revtex4-2}
\usepackage{xcolor}
\usepackage{parskip}
\usepackage{amsmath,amssymb,amsfonts,bm}
\usepackage{comment}
\usepackage{braket}
\usepackage[unicode=true,
            pdfusetitle,
            colorlinks=true,
            urlcolor=blue,
            anchorcolor=black,
            citecolor=red,
            filecolor=navyblue,
            linkcolor=red,
            menucolor=navyblue,
            linktocpage=true,
            pdfproducer=medialab]{hyperref}
\hypersetup{pdfauthor={Name}}
\DeclareSymbolFont{usualmathcal}{OMS}{cmsy}{m}{n}
\DeclareSymbolFontAlphabet{\mathcal}{usualmathcal}

\newcommand\SC[1]{\textcolor{black}{#1}}

\def\Psib{\overline{\Psi}}

\def\ha{{\bf\hat{a}}}
\def\hb{{\bf\hat{b}}}

\def\cT{{\cal T}}

\def\vn{{\vec{n}}}
\def\vm{{\vec{m}}}
\def\vM{{\vec{M}}}

\def\Psib{\overline{\Psi}}

\def\T{{\cal T}}

\begin{document}

\preprint{}

\title{Symmetries and Anomalies of Hamiltonian Staggered Fermions}

\author{Simon Catterall}
\email[]{smcatter@syr.edu}
\affiliation{Department of Physics, Syracuse University, Syracuse, 13244, New York, USA}

\author{Arnab Pradhan}
\email[]{arpradha@syr.edu}
\affiliation{Department of Physics, Syracuse University, Syracuse, 13244, New York, USA}

\author{Abhishek Samlodia}
\email[]{asamlodi@syr.edu}
\affiliation{Department of Physics, Syracuse University, Syracuse, 13244, New York, USA}

\date{\today}

\begin{abstract}
We review the shift (translation) and time reversal symmetries of Hamiltonian staggered fermions and their connection to continuum symmetries concentrating in particular on the case of massless fermions and (3+1) dimensions.
We construct operators using the staggered fields that implement these symmetries on finite lattices.  We show that shifts composed of an odd multiple of the elementary shift anti-commute with time reversal and are related to continuum axial transformations. We argue that the presence of these non-trivial commutation relations implies the existence of lattice 't Hooft anomalies. From the shifts we also construct a set of conserved, quantized charges that generate continuous
symmetries of the lattice theory. In general these do not commute with the vector
charge signaling further 't Hooft anomalies.
\end{abstract}

\maketitle

\section{Introduction}

In this paper, we focus on the symmetries of staggered fermions. Although
the symmetries of the Euclidean formulation, in which both time and space are discretized, are well known \cite{gliozzi1982spinor,vandenDoel:1983mf,Golterman:1984cy,SHARATCHANDRA1981205,KLUBERGSTERN1983447,Kilcup:1986dg}, 
the Hamiltonian
formalism developed in \cite{susskind} has received less attention in 
the case when the spatial dimension is greater than one~\footnote{Our work
has been influenced, however, by recent theoretical
work on staggered fermions in (1+1) dimensions given in 
\cite{Dempsey:2022nys, Chatterjee:2024gje}}.
In our work we focus on the structure of the shift and time reversal 
symmetries for Hamiltonian staggered fermions for arbitrary spatial dimension. 
We are particularly interested in understanding the connection between 
the anomalies seen in Euclidean formulations of staggered or K\"{a}hler-Dirac
fermions \cite{Butt:2021brl,Catterall:2022jky} and the structure of these theories as viewed from a Hamiltonian perspective. In particular, we would like to understand whether we can build
chiral lattice gauge theories by gauging certain discrete translation
symmetries of staggered fermions along the lines proposed in \cite{Catterall:2024jps}. These
discrete symmetries are called shift symmetries in the literature,
and, as we will discuss later,
can be thought of as a finite subgroup of the axial-flavor symmetry of the continuum theory. In particular our focus will be on understanding whether such
symmetries break in response to gauging other symmetries signaling the
presence of mixed 't Hooft anomalies - a phenomenon that
has has been observed
in other lattice systems 
\cite{Cheng:2022sgb,Seiberg:2023cdc,hersh}. In path integral approaches to quantum field theory, anomalies including mixed
anomalies, arise from a non-invariance of the fermion
measure.~\footnote{Notice that the usual ABJ anomaly can be thought of as a mixed anomaly since it corresponds to the breaking of a global axial symmetry in the presence of
background gauge field for a vector symmetry.} In contrast within the canonical formalism mixed anomalies
are realized when operators representing distinct 
symmetries do not commute. 
 
Following the procedure given in \cite{Seiberg:2023cdc} we construct explicit operators that implement the elementary
shifts $S_k$, time reversal ${\cal T}$
and a global $U(1)$ phase symmetry on a finite lattice. Our work can be seen as extension of recent work on Majorana chains 
in one spatial dimension \cite{Seiberg:2023cdc} and the Schwinger
model \cite{Dempsey:2022nys,Chatterjee:2024gje}. Motivation for our work can also be found in the phenomenon of
symmetric mass generation which requires the cancellation
of lattice 't Hooft anomalies 
\cite{Ayyar:2014eua,Ayyar:2016lxq,Catterall:2015zua,Butt:2021brl,Wang:2022ucy,Catterall:2022jky}
and the formulation of certain lattice chiral gauge theories 
using mirror fermions \cite{Zeng:2022grc,Catterall:2023nww}. Many of these studies start from the Hamiltonian formulation which provides a motivation for this work.

We start by reviewing the staggering procedure for the Hamiltonian formalism and then discuss the symmetries of the system focusing on the shift and time
reversal symmetries. We then construct finite operators that implement these
symmetries on the lattice and examine their commutator structure. We examine in particular
the case of three spatial dimensions showing the relationship of shift invariance to
continuum axial-flavor symmetries. Finally we are able to construct a series of exact continuous symmetries of the model by combining the phase symmetry with
the shift symmetries. These symmetries obey a non-trivial algebra which we conjecture encodes the continuum anomaly structure for vanishing lattice spacing.

\section{Hamiltonian staggered fermions}

Our starting point is the continuum Dirac Hamiltonian given by
\begin{equation}
    H = \int d^3x\,\Psib(x)\,\left(i\gamma_i\partial_i+m\right) \Psi(x)
\end{equation}
where $i$ are spatial indices running from $1\ldots d$. 
The lattice Hamiltonian 
is obtained by first introducing a cubic spatial lattice
and replacing the derivative with a symmetric finite difference
{\begin{equation}
    H = \sum_x \Psib(x)\left[i\gamma_i\,\Delta_i(x,y)+m\delta_{x,y}\right]\Psi(y)
\end{equation}
where $x$ now labels an integer position vector on the lattice with components $\{x_i\}$ and the symmetric difference operator is defined by
\begin{equation}
\Delta_i(x,y)=\frac{1}{2}\left(\delta_{y,x+i}-\delta_{y,x-i}\right)\end{equation}
We indicate a shift of the site $x$ by one lattice spacing
in the $i^{\rm th}$ direction by $(x+i)$. It is convenient to introduce the hermitian matrices $\alpha_i=\gamma_0\gamma_i$ and $\beta=\gamma_0$ 
and rewrite this as
\begin{equation}
    H = \sum_{x,y} \Psi^\dagger(x)\left[\,i\alpha_i\Delta_i(x,y)+m\beta\delta_{xy}\right]\Psi(y)
    \label{start}
\end{equation}
\SC{As is well known this lattice Hamiltonian suffers from fermion doubling
and describes $2^d$ degenerate Dirac fermions in the continuum limit. Staggered fermions
represent efforts to reduce this degeneracy.}
To obtain the staggered Hamiltonian, we first perform a unitary
transformation on $\Psi$ to a new basis $\chi$ as follows: 
\begin{align}
\label{unitary}
    \Psi(x) &= \alpha^x\chi(x)\nonumber\\
    \Psi^\dagger(x) &=  \chi^\dagger (x) (\alpha^x)^\dagger  
\end{align}
where 
\begin{equation}
    \alpha^x=\alpha_{1}^{x_1}\alpha_{2}^{x_2}\ldots \alpha_{d}^{x_d}
\end{equation}
The Hamiltonian in this basis is then given by
\begin{equation}
    H = \sum_{x,y}\chi^\dagger(x)\left[i\eta_i(x)\Delta_i(x,y)+m\epsilon(x)\delta_{xy}\beta\right]\chi(y)
    \label{hamiltonian}
\end{equation}
where $\eta_i(x) = (-1)^{x_1 + x_2 + ... + x_{i-1}}$ and
$\epsilon(x)=\left(-1\right)^{\sum_i x_i}$. Unlike the analogous situation
in Euclidean space the resultant operator for $m\ne 0$ is not proportional to the unit matrix
in spinor space. So we cannot stagger the
field by merely discarding all but one component of $\chi$
as one would do in that case. Instead, we can go back to eqn.~\ref{start} and
decompose $\Psi(x)$ into two components $\Psi_{\pm}(x)$ that
are eigenstates of $\beta$ 
\begin{equation}
    \Psi(x)=\Psi_+(x)+\Psi_-(x)
\end{equation}
where $P_\pm=\frac{1}{2}\left(1\pm\beta\right)$. 
The Hamiltonian is then
\begin{align}
    H=\sum_{x,y} &\left[\Psi^\dagger_+(x)\,i\alpha_i\Delta_i(x,y)\Psi_-(y)\SC{+}\right.\nonumber\\
    &\left.\Psi^\dagger_-(x)i\alpha_i\Delta_i(x,y)\Psi_+(y)\right]+\nonumber\\
    &m\sum_x \left(\Psi^\dagger_+(x)\Psi_+(x)-\Psi^\dagger_-(x)\Psi_-(x)\right)\label{Psi}
\end{align}
and write $\Psi_\pm$ as
\SC{
\begin{align}
    \Psi_\pm(x)&= P_\pm\alpha^x\chi(x)=\alpha^x\frac{1}{2}\left(1\pm\epsilon(x)\beta\right)\chi(x)\nonumber\\
    &=\alpha^x\frac{1}{2}\left(1\pm\epsilon(x)\beta\right)\left[\chi_+(x)+\chi_-(x)\right]\nonumber\\
    &=\alpha^x\frac{1}{2}\left[(1\pm\epsilon(x))\chi_+(x)+(1\mp\epsilon(x))\chi_-(x)\right]
\end{align}
Thus
\begin{align}
    \Psi_+=\alpha^x(\chi_{+e}+\chi_{-o})\nonumber\\
    \Psi_-=\alpha^x(\chi_{+o}+\chi_{-e})
\end{align}}
\SC{It is now possible to truncate the system by setting $\chi_{-o}=\chi_{-e}=0$. On substitution into eqn.~\ref{Psi} one obtains
\begin{align}\label{stagtrunc}
    H&=\sum_{x,i} \chi_{+e}^\dagger i\eta_i\Delta_i \chi_{+o}+\chi_{+o}^\dagger i\eta_i\Delta_i\chi_{+e}+\nonumber\\
    &m\left(\chi_{+e}^\dagger\chi_{+e}-\chi_{+o}^\dagger\chi_{+o}\right)
\end{align}
}
\SC{Since
the action is now diagonal in spinor indices one can further truncate $\chi_{+}(x)\equiv \chi(x)$ to a single component field and in this way obtain the final staggered fermion Hamiltonian:
\begin{equation}
H=\sum_{x,y,i}\chi^\dagger(x)i\eta_i(x)\Delta_i(x,y)\chi(y)+m\sum_x\epsilon(x)\chi^\dagger(x)\chi(x)
\end{equation}
This final step 
thins the degrees of freedom by another factor of two~\footnote{For simplicity we will continue to use the notation $\chi^\dagger$ going forward even though $\chi$ is a single component field. To show that this Hamiltonian is hermitian one must employ the result $(ab)^\dagger=b^\dagger a^\dagger$ for fermion operators.} and the resulting staggered Hamiltonian describes two Dirac fermions in three (spatial) dimensions \cite{susskind, Kogut:1974ag} and one Dirac fermion in one dimension in the continuum limit \cite{Chatterjee:2024gje,Dempsey:2022nys}. 
}
The canonical anti-commutators of the staggered fields are given by
\begin{equation}
    \{\chi^\dagger(x,t),\chi(x^\prime,t)\} = \delta_{x,x^\prime}
\end{equation}
with all other anti-commutators vanishing.
The equation of motion is  
\begin{align}
    i\frac{\partial\chi(x)}{\partial t}&=[H,\chi(x)]\nonumber\\
&=i\eta_i(x)\Delta_i(x,y)\chi(y)+m\epsilon(x)\chi(x)\label{eom}
\end{align}
It is not hard to verify that 
\begin{equation}
    \eta_i(x)\eta_j(x+i)+\eta_j(x)\eta_i(x+j)=2\delta_{ij}
\end{equation}
This result, together with the fact that the site parity operator $\epsilon(x)$
anticommutes with the symmetric difference operator $\Delta_i$,
ensures that the field $\chi$ satisfies a discrete Klein Gordon
equation
\begin{equation}
    \frac{\partial^2 \chi(x)}{\partial t^2}=\frac{1}{2}\sum_i \left[\chi(x+2i)+\chi(x-2i)-2\chi(x)\right]+m^2\chi(x)
\end{equation}
Notice the appearance of a discrete Laplacian operator on a block lattice with twice
the lattice spacing. This equation implies that there are $2^d$ 
degenerate solutions for every site on the block
lattice. These solutions can be constructed by performing
a translation accompanied by a phase shift $\xi_i(x)$ of the block lattice solution
for every site in the unit cell 
of the original lattice. These translations within the unit cell, or shifts as they are commonly called, play a crucial role in our
analysis and will be discussed later.

Let us try to understand how this works in more detail. For simplicity let us
restrict the following discussion to odd $d$.
The naive fermion on a d-dimensional spatial lattice
gives rise to $2^{d}$ Dirac fermions in the continuum limit because of doubling. After staggering one expects the continuum theory to correspond to $2^{\frac{d-1}{2}}$ Dirac fermions with a total of
$2^d$ complex components. These can be identified with the
staggered fields $\chi(x)$ 
living at the corners of the unit hypercube 
in the original lattice or equivalently the
unit cell in the block lattice. 
To expose the spin-flavor structure of the continuum fermion
we can build
a matrix fermion $\Lambda$ according to the rule
\begin{equation}
    \Lambda(x)=\frac{1}{8}\sum_{\{b\}}\chi(x+b)\alpha^{x+b}
    \label{matrix}
\end{equation}
where $\{b\}$ is a set of $2^d$ vectors with components $b_i=\{0,1\}$ corresponding to points in a unit cell. Clearly, as written, $\Lambda$ carries more degrees of freedom than the original staggered field. But the low momentum components \SC{can be taken as} independent and are enough to expose the
spin-flavor structure - the continuum spinors can then be read off from the columns of this matrix as
the lattice spacing is sent to zero \cite{Bock:1992yr}. \SC{As we will
show it also allows us to see the connection between shift symmetries
and the continuum axial-flavor symmetry}. A matrix mapping which
preserves the number of degrees of freedom is called the spin-taste basis and is given in terms of a distinct matrix 
field $\psi$ living on the block lattice only. This is reviewed in appendix C.

Finally, we note that
the Hamiltonian is also clearly invariant under a $U(1)$ phase symmetry in which $\chi(x)\to e^{i\theta}\chi(x)$ which will play an important role in our later
discussion. We now turn to a discussion of important additional symmetries of the
staggered Hamiltonian.

\section{Continuum flavor and lattice shift symmetries}
In this section we will focus on staggered fermions in $3+1$ dimensions but the arguments can be easily generalized to any dimension.
We start from a chiral basis
for the Dirac gamma matrices given by
\begin{equation}
    \gamma_\mu=\left(\begin{array}{cc}0&\sigma_\mu\\\bar{\sigma}_\mu&0\end{array}\right)
\end{equation}
where $\sigma_\mu=(I,\sigma_i)$ and $\bar{\sigma}_\mu=(I,-\sigma_i)$. 
In this case 
\begin{align}
    \alpha_i=\left(\begin{array}{cc}-\sigma_i&0\\0&\sigma_i\end{array}\right)\nonumber\\
\end{align}
For later reference we also list $i\alpha_j\alpha_k$ and $i\alpha_1\alpha_2\alpha_3$ in this basis:
\SC{
\begin{align}
i\alpha_j\alpha_k=\left(\begin{array}{cc}\epsilon_{jki}\sigma_i&0\\
0&\epsilon_{jki}\sigma_i\end{array}\right)\end{align}
\begin{align}
i\alpha_1\alpha_2\alpha_3=\left(\begin{array}{cc}
I&0\\0&-I\end{array}\right)
\end{align}
}
The matrix $\Psi$ in eqn.~\ref{matrix} then takes the form
\begin{equation}
    \Lambda=\left(\begin{array}{cc}\lambda_R&0\\0&\lambda_L\end{array}\right)
    \label{contferm}
\end{equation}
where
\begin{align}
\lambda_R & =\chi(x)I-\chi(x+i)\sigma_i-\frac{i}{2}\chi(x+i+j)\epsilon_{ijk}\sigma_k\nonumber\\
       & +iI\chi(x+\hat{1}+\hat{2}+\hat{3})\nonumber\\
\lambda_L & =\chi(x)I+\chi(x+i)\sigma_i-\frac{i}{2}\chi(x+i+j)\epsilon_{ijk}\sigma_k\nonumber\\
       & -iI\chi(x+\hat{1}+\hat{2}+\hat{3})
\end{align}
In the naive continuum limit the massless staggered field thus gives
rise to doublets of left
and right-handed Weyl fields transforming under two independent $SU(2)$  symmetries. These 
symmetries protect the continuum theory 
from developing bilinear mass terms~\footnote{A Majorana mass for either left or right handed doublet vanishes identically }. In addition the theory is invariant under both the vector $U(1)$ symmetry discussed
earlier and a
singlet axial symmetry $U(1)_A$ in which the $\lambda_R$ and $\lambda_L$ carry
opposite charges.  Of course the crucial question is whether sufficient
lattice symmetries exist that guarantee that these continuum symmetries
emerge as the lattice spacing is sent to zero. Part of the answer
lies in the existence of exact translation-by-one or {\it shift} symmetries
of the lattice Hamiltonian. We now turn to these
symmetries and their relation to continuum symmetries.

The continuum symmetries act by right multiplication of a continuum
matrix fermion given by eqn.~\ref{matrix} as $a\to 0$ by an axial-flavor
transformation matrix $F$. This takes the form $F=e^{i\theta_A\alpha_A}$ where the hermitian basis $\alpha_A$ is given
in terms of products of the individual $\alpha_i$ matrices:
\begin{equation}
\alpha_A=\{\alpha_i,i\alpha_i\alpha_j, i\alpha_1\alpha_2\alpha_3\}\quad {\rm where}\; i=1\ldots 3\label{shifts}
\end{equation}
It can be seen that this yields
the continuum flavor group $SU(2)\times SU(2)\times U_A(1)$ as described above.

Staggered fermions, being discretizations of K\"{a}hler-Dirac fermions \cite{Catterall:2022jky}, are invariant under a twisted
rotation group corresponding to the diagonal subgroup of the flavor and rotation symmetries  \cite{kahler,landau,Kilcup:1986dg,Golterman:1984cy,
Catterall:2024jps}. Upon discretization, the rotational symmetries
are restricted to the cubic group and
hence the flavor rotation angles are similarly restricted to odd multiples
of $\frac{\pi}{2}$. Then the remaining
elementary discrete flavor rotation acting on the continuum matrix field becomes (see \cite{Bock:1992yr})
\begin{equation}
    \Lambda(x)\to \Lambda(x) e^{i\frac{\pi}{2}\alpha_j}=\Lambda(x)\,i\alpha_j
\end{equation}
The lattice equivalent is
\begin{equation}
    \Lambda(x)\to \Lambda(x+j)\,i\alpha_j
\end{equation}

We will now show that this induces a unit translation or {\bf shift}
on the staggered fields. Acting on the lattice fermion matrix given in
eqn.~\ref{matrix}, the transformed matrix is given by
\begin{align}
    \Lambda^{'}(x)&=\sum_{\{b\}}\chi(x+b+j)\alpha^{x+b}i\alpha_j\nonumber\\
    &=i\sum_{\{b\}}\xi_j(x+b)\chi(x+b+j)\alpha^{x+b+\textcolor{red}{j}}
\end{align}
where one must anticommute the $\alpha_j$ matrix from the right which produces the phase factor $\xi_j(x+b)$ \SC{where $\xi_j(x)=\left(-1\right)^{\sum_{k=j+1}^d x_k}$}.
The net effect is clearly to just produce an elementary
shift 
\begin{align}
\chi(x)&\stackrel{S_j}{\to} i\xi_j(x)\chi(x+j)\nonumber\\
\chi^\dagger(x)&\stackrel{S_j}{\to} -i\xi_j(x)\chi^\dagger(x+j)
\end{align}
Thus there is a direct connection between the continuum transformation $e^{i\frac{\pi}{2}\alpha_j}$ and the elementary shift $S_j$.

It is straightforward to examine 
the invariance of the Hamiltonian under such a shift $S_k$
\begin{align}
    H\to &\sum_{x,i} i\eta_i(x)\chi^\dagger(x+k)\xi_k(x)\xi_k(x+i)\frac{1}{2}\left[\chi(x+i+k)-\right.\nonumber\\
    &\left.\quad\chi(x-i+k)\right]+\sum_x m\epsilon(x)\xi_k(x)^2\chi^\dagger(x+k)\chi(x+k)\nonumber\\
    = &\sum_{x,i}i\eta_i(x-k)\chi^\dagger(x)\xi_k(x-k)\xi_k(x+i-k)\frac{1}{2}\left[\chi(x+i)-\right.\nonumber\\
    &\left.\quad\chi(x-i)\right]+\sum_x m\epsilon(x-k)\chi^\dagger(x)\chi(x)\nonumber\\
    = &\sum_x i\eta(x)\chi^\dagger(x)\frac{1}{2}\left[\chi(x+i)-\chi(x-i)\right]\nonumber\\
    &-\sum_x m\epsilon(x)\chi^\dagger(x)\chi(x)
\end{align}
where we have used the result $\eta_i(k)\xi_k(i)=1$. Notice that the Hamiltonian
is only invariant under $S_k$ if $m=0$.  
It should be clear
that elementary shifts can be applied consecutively to yield additional
symmetries. For example
the double shift $S_{ij}$:
\begin{equation}
    \chi(x)\stackrel{S_{ij}}{\to} S_i S_j\, \chi(x)=-\xi_j(x)\xi_i(x+j)\chi(x+i+j)\quad i\ne j
\end{equation}
It is trivial to see that $S_i S_j=-S_j S_i$. Thus, just as $S_k$ is
associated with the continuum symmetry generator $\alpha_k$ the properties of the
double shift $S_{ij}$ mimic those of the generator $i\alpha_i\alpha_j$. It is
important to notice that in the case of the double
shift the Hamiltonian
is invariant even for non-zero mass. This suggests that
even lattice shifts are associated with vector transformations in the continuum theory. If one goes to a chiral basis in the continuum it is easy to verify that the generators $\alpha_i\alpha_j$ indeed act as vector symmetries.

Conversely the Hamiltonian is only invariant under an odd number of shifts if the mass is
zero suggesting odd shifts are associated with continuum axial
transformations. This can also be explicitly verified by going
to a chiral basis. An explicit example of this is the 
triple shift $S_1S_2S_3$ which corresponds to
the generator of the singlet axial $U_A(1)$ symmetry $\gamma_5=i\alpha_1\alpha_2\alpha_3$.

It is important to recognize that the
connection between lattice shifts and continuum flavor is only 
one to one up to ordinary translations. For example,
a double shift along the same direction yields a simple translation $T$ on the block lattice:
\SC{\begin{align}
    \chi(x)&\stackrel{S_i^2}{\to}-\xi_i(x)\xi_i(x+i)\chi(x+2i)\nonumber\\&=-\chi(x+2i)=-T_i\left[\chi(x)\right]
\end{align}}
Similarly,
performing a $S_{ij}$ shift followed by a $S_{j}$ shift yields
\begin{align}
S_{ij}S_j \chi(x)&\to S_{ij}i\xi_j(x)\chi(x+j)\nonumber\\
&= -i\xi_{j}(x+j)\xi_i(x+2j)\xi_j(x)\chi(x+2j+i)\nonumber\\
&=-i\xi_i(x)T_j\chi(x+i)\nonumber\\
&=-T_jS_i \chi(x)
\end{align}
or more generally $[S_{ij},S_j]=-2T_jS_i$.
That is, the combination of two shifts generates another shift up to a block translation. In a similar fashion the double
shifts satisfy the relation
\begin{equation}
    [S_{ij},S_{jk}]=2S_{ik}T_j
\end{equation}
The shift symmetries hold even in the presence of gauge interactions provided the gauge link field $U_i(x)$ transforms similarly under shifts
\begin{equation}
    U_i(x)\stackrel{S_j}{\to} U_i(x+j)
\end{equation}
It should now be clear that the staggered fermion shift symmetries form a discrete subgroup of the continuum symmetries corresponding to discrete axial-flavor
transformations~\footnote{The word flavor is often called taste in the lattice gauge theory literature.}. This group comprises the elements given in eqn.~\ref{shifts} together with their negatives and the identity. 
When the lattice mass is non-zero only the even shifts are good lattice
symmetries. These form a discrete subgroup of the $SU(2)_V$ symmetry of the
continuum theory. \SC{In section \ref{renorm} we show that the exact symmetries
of the lattice theory are sufficient to protect the theory against
developing relevant mass terms and these continuum
symmetries are hence restored automatically
as the lattice spacing is sent to zero.}
These conclusions parallel similar arguments that can be made in the Euclidean theory - see \cite{Golterman:1984cy,KLUBERGSTERN1983447,vandenDoel:1983mf,gliozzi1982spinor,SHARATCHANDRA1981205} 
for a discussion of these issues in the context of the Euclidean theory.

\section{Time reversal on and off the lattice}
In (3+1) dimensions the discrete symmetries
of charge conjugation $C$ and time reversal $T$ (where $t\to -t$) act on a continuum spinor field $\Psi$ in the following manner:
\begin{equation}
    \Psi\stackrel{C}{\to} \gamma_2\Psi^*=\beta\alpha_2\Psi^*\\
\end{equation}
and
\begin{align}
    \Psi&\stackrel{T}{\to}\gamma_1\gamma_3\Psi=\SC{-}\alpha_1\alpha_3\Psi\nonumber\\
    i&\stackrel{T}{\to}-i
\end{align}
In particular the combination $\cT=CT$ acts as follows
\begin{align}
    \Psi&\stackrel{\cT}{\to}\Gamma\Psi^*\nonumber\\
    i&\stackrel{\cT}{\to}-i
\end{align}
where $\Gamma=\beta\alpha_2\alpha_1\alpha_3$. A standard calculation shows
that if $\Psi(x,t)$ is a solution of the EOM then so is $\Gamma\Psi^*(-t,x)$.
Performing the $CT$ transformation on the Hamiltonian after the unitary
transformation given in eqn.~\ref{unitary} shows that the symmetry operation $\cT$ acts on staggered fields as follows
\begin{align}
    \chi(x)&\stackrel{\cT}{\to} \epsilon(x)\chi^*(x)=\epsilon(x)(\chi^\dagger(x))^T\\\nonumber
    i&\stackrel{\cT}{\to}-i
\end{align}
\SC{
Under ${\cal T}$ the kinetic term $K\to K^\prime$ 
\begin{align}
    K^\prime&=\sum_x \chi(x)\epsilon(x)[\,-i\eta_i(x)\Delta_i(x,y)]\epsilon(y)\chi^*(y)\nonumber\\
    &=\sum_x \chi(x)\,i\eta_i(x)\Delta_i(x,y)\chi^*(y)\nonumber\\
    &=\sum_x -\chi^*(y)\,i\eta_i(x)\Delta_i(x,y)\chi(x)\nonumber\\
    &=K
\end{align}
where the last line follows from the fact that $\Delta_i$ is an antisymmetric
matrix and $\chi, \chi^*$ anticommute. Thus the kinetic operator is time reversal invariant.
In contrast, it is easy to see that
the mass term $\sum_x\epsilon(x)\chi^*(x)\chi(x)$ is not invariant under ${\cal T}$.}
For brevity we will refer to this
product of charge conjugation and time reversal symmetry as simply time reversal $\cT$ in the rest of the paper~\footnote{Clearly $\cT^2=1$.}.
At this point it is important to notice that the elementary shift symmetry
does {\bf not} commute with time reversal:
\begin{align}
    \chi(x)&\stackrel{\cT}{\to} \epsilon(x)\chi^*(x)\nonumber\\
    \chi(x)&\stackrel{S_k\cT}{\to} -i\epsilon(x)\xi_k(x)\chi^*(x+k)\nonumber\\
    \chi(x)&\stackrel{S_k}{\to} i\xi_k(x)\chi(x+k)\nonumber\\
    \chi(x)&\stackrel{\cT S_k}{\to} -i\xi_k(x)\epsilon(x+k)\chi^*(x+k)=i\epsilon(x)\xi_k(x)\chi^*(x+k)
\end{align} Clearly, the
two symmetries anti-commute.~\footnote{Notice that we can multiply the elementary shift symmetry by an arbitrary phase $\alpha$. But this does not change the anti-commutation property with ${\cal T}$.} Indeed, this statement
is true for any shift composed of an odd number of elementary shifts.

\section{Renormalization}
\label{renorm}
Let us summarize our conclusions so far. The staggered Hamiltonian with
zero mass is invariant
under a $U(1)$ phase symmetry, time reversal and a set of 
shift symmetries that form a discrete subgroup of 
the continuum axial-flavor group. The lattice theory is also invariant under discrete rotations and axis inversion as described in appendix C 
and a charge conjugation symmetry $C$:
\begin{equation}
    \chi(x)\to (\chi^\dagger(x))^T\quad \chi^\dagger(x)\to \chi^T(x)
\end{equation}

\SC{To understand
whether these symmetries enhance to yield a Lorentz and time reversal 
invariant theory equipped with the full $SU(2)\times SU(2)\times U(1)$ continuum axial-flavor
symmetry we need to write down all relevant and marginal lattice operators that are invariant under the lattice symmetries and determine
whether any of these correspond to relevant (or marginal) operators that break the continuum
symmetries.
The only candidate terms one can construct correspond to fermion bilinears coupling sites within the unit cell of the lattice. 
The only operators that are both shift and $U(1)$ invariant correspond
to fermion fields connected by a string of $\eta_i(x)$ link phases along a 
path between $x$ and $x+\vec{n}$:
\begin{equation}
    \chi^\dagger(x)\left[\prod_{n_i\ne 0}\eta_{n_i}(x+\sum_{j<i}n_j)\right]\chi(x+\vec{n})+{\rm h.c}
\end{equation}
We have already shown the shift invariance of the kinetic term which takes
this form. As another example consider the following operator
\begin{equation}
    \sum_x\chi^\dagger(x)i\eta_i(x)\eta_j(x+i)\chi(x+i+j)\quad i\ne j
\end{equation}
Under a shift $S_k$ it becomes
\begin{align}
&\sum_x \chi^\dagger(x+k)\xi_k(x)i\eta_i(x)\eta_j(x+i)\xi_k(x+i+j)\times\nonumber\\&\qquad\chi(x+i+j+k)\nonumber\\
&=\sum_x \chi^\dagger(x) i\eta_i(x)\eta_i(k)\eta_j(x+i)\eta_j(k)\xi_k(i)\xi_k(j)\chi(x+i+j)\nonumber\\
&=\sum_x \chi^\dagger(x)i\eta_i(x)\eta_j(x+i)\chi(x+i+j)
\end{align}
where we have used the result $\eta_i(k)\xi_k(i)=1$ twice. It should be clear that a similar result will be obtained for any string of $\eta$-link phases terminated by fermion fields.}
\SC{
Notice that any insertion of $\xi_k(x)$ or $\epsilon(x)$ into
this expression will break the shift symmetry. For example the term
\begin{equation}
    \sum_x \chi^\dagger(x)\xi_i(x)\chi(x+i)
\end{equation} transforms to
\begin{align}
    &\sum_x \chi^\dagger(x+k)\xi_k(x)\xi_i(x)\xi_k(x+i)\chi(x+i+k)\nonumber\\
    &=\sum_x \chi^\dagger(x)\xi_i(x)\chi(x+i)\left[\xi_k(i)\xi_i(k)\right]
\end{align}
The factor in square brackets is $2\delta_{ik}-1$.}

\SC{
The goal of this section is to write down all such terms and then decide whether they are invariant under both ${\cal T}$ and $C$. We will do this systematically according to the number of non-zero elements or links in the
vector $\vec{n}$.}
\SC{\subsection{Zero link operators}
There are just two hermitian operators of this kind
\begin{enumerate}
\item $\epsilon(x)\chi^\dagger(x)\chi(x)$
\item $\chi^\dagger(x)\chi(x)$
\end{enumerate}
Both lead to relevant mass terms in the continuum limit.
We have already shown that the former is not
invariant under single shifts or time reversal. It is also
not invariant under $C$. While the second operator 
is invariant under shifts it is easy to see
that it is not invariant under ${\cal T}$ or $C$ since $\{\chi(x),\chi^\dagger(y)\}=\delta_{xy}$
\subsection{One link operators}
The following hermitian lattice operators are possible
\begin{enumerate}
    \item $\chi^\dagger(x) i\eta_i(x)\left[\chi(x+i)-\chi(x-i)\right]$
    \item $\chi^\dagger(x) \eta_i(x)\left[\chi(x+i)+\chi(x-i)\right]$
\end{enumerate}
These expressions should be summed over the index $i$ to enforce rotational
invariance but we will suppress this aspect for simplicity.
The first one is just the original kinetic term and we have shown that
it is invariant under ${\cal T}$ and $C$. 
The second is a mass term. Under time reversal it becomes
\begin{align}
 &-\sum_x \epsilon(x)\chi^T(x)\eta_i(x)\epsilon(x)\left[(\chi^\dagger(x+i))^T+(\chi^\dagger(x-i))^T\right] \nonumber\\
 &=\sum_x \left[(\chi^\dagger(x+i))^T+(\chi^\dagger(x-i))^T\right]\eta_i(x)\chi(x)\nonumber\\
 &=\sum_x \chi^\dagger(x)\eta_i(x)\left[\chi(x+i)+\chi(x-i)\right]
\end{align}
It is thus ${\cal T}$ invariant. Under $C$ it becomes
\begin{align}
    &\sum_x \chi^T(x)\eta_i(x)(\chi^\dagger(x+i)+\chi^\dagger(x-i))^T\nonumber\\
    &=\sum_x -\chi^\dagger(x)\eta_i(x)\left[\chi(x+i)+\chi(x-i)\right]
\end{align}
It is hence not $C$ invariant.
\subsection{Two link operators}
We consider
%\begin{enumerate}
    %
    \begin{equation}\chi^\dagger(x)i\eta_i(x)\eta_j(x+i)\left[\chi(x+i+j)+\chi(x-i-j)\right]\end{equation}
%\end{enumerate}
In principle to enforce rotational invariance this expression should be
summed over all sets of neighbor points $x\pm i\pm j$ but again we will ignore this requirement for the purpose of
testing ${\cal T}$ and $C$ invariance other than requiring that the term be hermitian. Notice that if we left off the factor of $i$ we could seemingly construct term involving $\left[\chi(x+i+j)-\chi(x-i-j)\right]$.
However this clearly gives rise to a second derivative term in the
continuum and so we neglect such irrelevant operators in our analysis.
The first half of this term transforms under ${\cal T}$ as
\begin{align}
    &\sum_x -i\epsilon(x)\chi^T(x)\eta_i(x)\eta_j(x+i)\epsilon(x+i+j)(\chi^\dagger(x+i+j))^T\nonumber\\
    &=\sum_x i\chi^\dagger(x+i+j)\eta_i(x)\eta_j(x+i)\chi(x)\nonumber\\
    &=\sum_x i\chi^\dagger(x)\eta_i(x)\eta_j(x+i)\chi(x-i-j)\left[\eta_i(j)\eta_j(i)\right]
\end{align}
But $\left[\eta_i(j)\eta_j(i)\right]=-1$ for $i\ne j$ and so this term
violates ${\cal T}$. It can be shown to be invariant under $C$.
\subsection{Three link operators}
We consider the following mass-like operator
\begin{equation}
    \chi^\dagger(x)i\eta_i(x)\eta_j(x+i)\eta_k(x+i+j)\chi(x+i+j+k)+{\rm h.c}
\end{equation}
Following the same strategy as above it can be shown to be ${\cal T}$ invariant but not $C$ invariant.}

\SC{
Let us summarize our conclusions. The original massless Hamiltonian is invariant under $U(1)$, shift, ${\cal T}$ and $C$ symmetries. The only marginal or relevant fermion bilinear constructed from fields
within the unit cell that is invariant under all these symmetries is
the original kinetic operator. In our arguments
we have left out any gauge field. However it is straightforward to extend
the analysis to gauged fermion bilinears. The transformation of any gauge field under the shift 
and $U(1)$ symmetries has already been discussed. Under both $C$ and the ${\cal T}$ symmetry the gauge field is taken to transform as
\begin{equation}
    U_i(x)\to U_i^*(x)
\end{equation}
One can then repeat the previous analysis with the same conclusion: that
the only operator in the massless theory that remains invariant
under all the lattice symmetries in the presence of gauge fields remains
the kinetic term. Thus the theory does not suffer from
additive mass renormalization and the continuum symmetries should be restored
automatically as the lattice spacing is sent to zero. There is one caveat - 
clearly the coupling to the kinetic operator can be renormalized due
to interactions. This would correspond to a renormalization of
the speed of light. 
}

\section{Real fields and symmetry operators}
To derive explicit operators that
implement shifts and time reversal on a finite lattice
it is useful to first re-express the staggered fermion $\chi$ in terms
of real fields $\lambda^1$ and $\lambda^2$:
\begin{align}
    \chi(x)&=\frac{1}{2}\left(\lambda^1(x)+i\lambda^2(x)\right)\nonumber\\
\chi^\dagger(x)&=\frac{1}{2}\left(\lambda^1(x)-i\lambda^2(x)\right)\nonumber\\
\end{align}
The massless Hamiltonian is then 
\begin{equation}
    H=\frac{1}{4}\sum_{x,j}\sum_{a=1}^2 \lambda^a(x)\,i\eta_j\textcolor{red}{(x)}\Delta_j\,\lambda^a(x)
\end{equation}
and the equal time \SC{anti-commutators} become
\begin{equation}
    \{\lambda^a(x),\lambda^b(x^\prime)\}=2\delta^{ab}\delta(x,x^\prime)
\end{equation}
The $U(1)$ symmetry discussed earlier
yields an $SO(2)$ symmetry acting on the real fields
\begin{equation}
\lambda^a(x)=e^{\theta R_{ab}}\lambda^b(x)
\end{equation}
where $R^{ab}=\epsilon^{ab}$.
Similarly, time reversal ${\cal T}$ acts as
\begin{align}
    \lambda^1(x)&\stackrel{{\cal T}}{\to}\epsilon(x)\lambda^1(x)\nonumber\\
    \lambda^2(x)&\stackrel{\cT}{\to} \epsilon(x)\lambda^2(x)
\end{align} 
We can write the elementary shift symmetry described in the previous section
as $S_k=R\hat{S}_k$
where
$\hat{S}_k$ is given by
\begin{equation}
    \lambda^a(x)\stackrel{\hat{S}_k}{\to} \xi_k(x)\lambda^a(x+k)
\end{equation}
In fact, it should be clear that the massless Hamiltonian is actually invariant under two separate half shifts given by
\begin{align}
    \lambda^1(x)&\stackrel{A_k}{\to} \xi_k(x)\lambda^1(x+k)\nonumber\\
    \lambda^2(x)&\stackrel{A_k}{\to} \lambda^2(x)
\end{align}
and 
\begin{align}
    \lambda^2(x)&\stackrel{B_k}{\to} \xi_k(x)\lambda^2(x+k)\nonumber\\
    \lambda^1(x)&\stackrel{B_k}{\to} \lambda^1(x)
\end{align}
with
\begin{equation}
    \hat{S}_k=A_k\,B_k
\end{equation}
The $B$ half shift is precisely the same symmetry considered in \cite{Chatterjee:2024gje} for a (1+1) dimensional staggered fermion model.

Let us now construct operators that implement $A_k$, $B_k$ and $R$
on a finite lattice equipped with
periodic boundary conditions.
As a warm up let us start with a one dimensional
lattice with $L$ sites and coordinate $x\equiv x_1=0\ldots L-1$. For staggered
fermions $L$ must be even and for $d=1$ the phase $\xi_1(x)=1$.
The $A_1$ shift
can then be implemented by the action of the shift operator 
\begin{equation}
   \lambda^{1,2}\to A_{1}^{-1} \lambda^{1,2} A_{1}. 
\end{equation}
where
\begin{equation}
A_1 = 2^{-L/2}\prod^{L-1}_{x=0}\Big(1-\lambda^1(x)\lambda^1(x+1)\Big),
\end{equation}
and  
\begin{equation}
    A_1^{-1}=2^{-L/2}\prod_{x=0}^{L-1}\Big(1+\lambda^1(x)\lambda^1(x+1)\Big),
\end{equation}
To see this one uses the results 
\begin{align}
    -\lambda^1(x+1)&=\frac{1}{2}\left[1+\lambda^1(x)\lambda^1(x+1)\right]\lambda^1(x)\left[1-\right.\nonumber\\
    &\left.\quad\lambda^1(x)\lambda^1(x+1)\right]\nonumber\\
    \lambda^2(x)&=\frac{1}{2}\left[1+\lambda^1(x)\lambda^1(x+1)\right]\lambda^2(x)\left[1-\right.\nonumber\\
    &\left.\quad\lambda^1(x)\lambda^1(x+1)\right]\nonumber\\
    1&=\frac{1}{2}\left[1+\lambda^1(x)\lambda^1(x+1)\right]\left[1-\right.\nonumber\\
    &\left.\quad\lambda^1(x)\lambda^1(x+1)\right]
\end{align}
A similar result follows for $B_1$ which is given by
\begin{equation}
B_1 = 2^{-L/2}\prod^{L-1}_{x=0}\Big(1-\lambda^2(x)\lambda^2(x+1)\Big),
\end{equation}
Combining the $A$ and $B$ shifts one obtains
\begin{equation}
\hat{S}_1=2^{-L}\prod_{a=1}^2\prod^{L-1}_{x=0}\Big(1-\lambda^a(x)\lambda^a(x+1)\Big)
\end{equation}
The time reversal operator can also be implemented in a similar way. When $L$ is a multiple of 4, it can be achieved using the operator ${\cal T}$:
\begin{equation}
    {\cal T} = {\cal K}\,\bigg(\prod_{x\;{\rm even}}\lambda^{1}(x)\bigg)\bigg(\prod_{x\; {\rm odd}}\lambda^{2}(x)\bigg).    
\end{equation}
where ${\cal K}$ represents complex conjugation.
When $L$ is an odd multiple of 2 this gives the wrong site parity, so we use $G{\cal T}$ instead, where $G$ is the fermion parity operator
\begin{equation}
    G\lambda^a(x)G^{-1} = -\lambda^a(x).
\end{equation}
with $G=\prod_x\prod_{a=1}^2 \lambda^a(x)$.

For a two dimensional lattice with
coordinates $(x_1,x_2)$ with $x_i=0\ldots L-1$ the story is similar.
An A-shift along $x_1$ is given by the action of a shift operator $A_1$ 
\begin{align}
A_1 = 2^{-L^2/2}\prod_{x_2=0}^{L-1}&\prod^{L-1}_{x_1=0}\nonumber\\
&\Big(1-\xi_1(x)\lambda^{1}(x_1,x_2)\lambda^{1}(x_1+1,x_2)\Big),
\end{align}
with $\xi_1(x)=\left(-1\right)^{x_2}$. 
Similarly, a shift along $x_2$ is generated by the operator 
\begin{align}
A_2 = 2^{-L^2/2}\prod_{x_1=0}^{L-1}&\prod^{L-1}_{x_2=0}\nonumber\\
&\Big(1-\xi_2(x)\lambda^{1}(x_1,x_2)\lambda^{1}(x_1,x_2+1)\Big).
\end{align}
with $\xi_2(x)=1$.
The B-shifts work in the same way with $\lambda^1(x)\to \lambda^2(x)$. This
allows us to write $\hat{S}_k$ in the form
\begin{align}
  \hat{S}_1 &= 2^{-L^2}\prod_{a=1}^2\prod_{x_2=0}^{L-1}\prod^{L-1}_{x_1=0}\Big(1-\xi_1(x)\lambda^{a}(x_1,x_2)\lambda^{a}(x_1+1,x_2)\Big)\nonumber\\
  \hat{S}_2 &= 2^{-L^2}\prod_{a=1}^2\prod_{x_1=0}^{L-1}\prod^{L-1}_{x_2=0}\Big(1-\xi_2(x)\lambda^{a}(x_1,x_2)\lambda^{a}(x_1,x_2+1)\Big)
\end{align}
When $L$ is a multiple of 4 time reversal can be achieved using
\begin{equation}
    {\cal T} = {\cal K}\,\bigg(\prod_{x\;{\rm even}}\lambda^{1}(x_1,x_2)\bigg)\bigg( \prod_{x\; {\rm odd}}\lambda^{2}(x_1,x_2)\bigg),   
\end{equation}
while when $L$ is an odd multiple of 2 we again use $G{\cal T}$ instead.
In three dimensions the $\hat{S}$ shifts are
\begin{align}
    \hat{S}_1 &= 2^{-L^3}\prod_{a=1}^2\prod_{x_3=0}^{L-1}\prod_{x_2=0}^{L-1}\prod^{L-1}_{x_1=0}\nonumber\\
    &\Big(1-\xi_1(x)\lambda^{a}(x_1,x_2,x_3)\lambda^{a}(x_1+1,x_2,x_3)\Big)\nonumber\\
    \hat{S}_2 &= 2^{-L^3}\prod_{a=1}^2\prod_{x_1=0}^{L-1}\prod_{x_3=0}^{L-1}\prod^{L-1}_{x_2=0}\nonumber\\
    & \Big(1-\xi_2(x)\lambda^{a}(x_1,x_2,x_3)\lambda^{a}(x_1,x_2+1,x_3)\Big)\nonumber\\ 
    \hat{S}_3 &=2^{-L^3}\prod_{a=1}^2\prod_{x_2=0}^{L-1}\prod_{x_1=0}^{L-1}\prod^{L-1}_{x_3=0}\nonumber\\
    &\Big(1-\xi_3(x)\lambda^{a}(x_1,x_2,x_3)\lambda^{a}(x_1,x_2,x_3+1)\Big) 
\end{align}
Time reversal when $L=0\;{\rm mod}\;4$ is given by
\begin{align}
    {\cal T} = {\cal K}\,\bigg(\prod_{\sum_i x_i = {\rm even}}\lambda^{1}(x_1,x_2,x_3)\bigg)\bigg( \prod_{\sum_i x_i = {\rm odd}}\lambda^{2}(x_1,x_2,x_3)\bigg)
\end{align}
with the same modification as before for $L=0\;{\rm mod}\;2$.
We can also write down an operator in terms of the fermion fields that implements the $R$ operation. It is given by
\begin{equation}
    \hat{R}=2^{-L^d}\prod^{L^d}_{x=0}\Big(1-\lambda^1(x)\lambda^2(x)\Big)\label{vector}
\end{equation}
To write down operators that correspond
to multiple shifts one simply compounds a series of single shift operators as
discussed earlier. 
As observed earlier, \SC{one can
verify that the odd shift operators anti-commute with $\T$.}
\begin{equation}
    \{S_k,{\cal T} \}=0
\end{equation}
Furthermore, the definition of $S_k$ involves an element $R$ of the $U(1)$
symmetry. Notice that $R$ satisfies $R^4=1$ and hence $R$ belongs to a $Z_4$
subgroup. The shifts are clearly symmetries for any element of this $Z_4$.
The fact that $S_k$ and $\cT$ do not commute implies that any attempt to gauge this $Z_4$ subgroup will break $\T$ - a mixed lattice 't Hooft anomaly. In particular,
the singlet $U_A(1)$ symmetry corresponding to
the 3-shift $S_{123}$ will be broken if this subgroup of the vector
symmetry is gauged. This is analogous to the result obtained
in \cite{Chatterjee:2024gje}.
In the next section we will explore how these 't Hooft anomalies can
be canceled.

\section{Anomaly cancellation}

The question we would like to address is whether the mixed 't Hooft anomaly
we found can be canceled. One way to approach this question is to ask whether
we can design interactions that can gap the fermions without breaking the $Z_4$
symmetry. One way to do the latter is to add $Z_4$ invariant four fermion
interactions to the theory. The simplest term we can write down takes
the form~\footnote{This is the operator that has been the best studied both analytically and numerically in the literature but clearly there are many others corresponding to displacing some of the fields within the unit cell of the lattice. }
\begin{equation}
    G\sum_x \chi^1(x)\chi^2(x)\chi^3(x)\chi^4(x)+{\rm h.c}
\end{equation}
which requires four complex staggered
fermions and exhibits an explicit $SU(4)=SO(6)$ global symmetry as well
as time reversal and shift invariance. 
For $G\to 0$ one expects the ground state to be eight fold
degenerate since it corresponds to eight non-interacting real
staggered fermions. However, for $G\to\infty$ the ground state
is given by diagonalizing the single site Hamiltonian. It was shown in \cite{Poppitz:2010at} and \cite{You:2014vea} that the ground state 
of this system is in fact a singlet. Indeed, in the 
latter paper it was shown how to construct a variety of four
fermion terms with differing symmetry groups (the maximal symmetry being $SO(7)$)
that result in a non-degenerate ground state - generalizing the original result of Kitaev et al \cite{kitaev}. The singlet nature of the ground state
implies that the system is incapable of undergoing spontaneous symmetry breaking.
This fact in turn implies that the system is free of 't Hooft anomalies.
This phenomenon of producing a gapped, invariant ground state 
has been termed symmetric mass generation and has already been observed
in staggered fermion models with four fermion interactions \cite{kitaev,Wang:2022ucy,Ayyar:2014eua,Ayyar:2015lrd,Butt:2021koj,Catterall:2016dzf,Catterall:2015zua,Butt:2018nkn}. 
Clearly the model can be gapped in this way for eight real staggered fermions each of which yields two Majorana fermions in the naive continuum limit. Thus this analysis confirms the $Z_{16}$ anomaly cancellation condition for Majorana
fermions in (3+1) dimensions \cite{Razamat:2020kyf,Wang:2020iqc,You:2014vea,Garcia-Etxebarria:2018ajm}.

One might be tempted to ask whether this anomaly cancellation
condition can be obtained directly from the algebra of operators. Consider $N$ flavors of massless staggered fermion.
The operators needed to implement shifts and time reversal take the form of products of mutually commuting terms for each flavor 
eg
\begin{equation}
    S_k=\prod_{a=1}^N S_k^{a}\quad \cT=\prod_{a=1}^N \cT^{a}
\end{equation}
where $S_k^{a}$ and $\cT^a$ denote the operators for a single
flavor derived in the previous section. Even though $\{S_k^a,\cT^a\}=0$ it is easy to verify that $[S_k,\cT]=0$ for $N=2k$. Thus, it appears in $d=3$ one would need four Dirac or eight Majorana fermions to cancel this mixed anomaly. In contrast,
the gapping argument tells us SMG should only be possible for
$N=4k$ i.e sixteen Majorana fermions in $d=3$. This discrepancy between the number of fermions needed to gap the system
and the number needed to cancel off a naive 't Hooft anomaly has been noted previously in the literature - see eg \cite{Witten_2016}. The $Z_{16}$ anomaly cancellation condition for the discrete
spin-$Z_4$ symmetry of continuum Weyl fields can not be seen by considering chiral fermions
on the torus where only a $Z_8$ condition is found. One must instead consider fermions propagating on manifolds with different topology to see the $Z_{16}$ classification.

\section{Conserved charges and continuous symmetries}
%\color{blue}
We have seen that the theory admits a phase (vector) symmetry
that manifests as an $SO(2)$ rotation on the doublet of real fermions $\lambda(x)$ at each site. 
The explicit expression for the generator of this second quantized symmetry operator was given in eqn.~\ref{vector}. In fact an arbitrary $SO(2)$ rotation can be generated using the operator $U_V=e^{-i\theta Q_V}$
whose action on the doublet field $\lambda$ is given by
\begin{equation}
    U_V\lambda(x)U_V^\dagger
\end{equation}
where the vector charge is given by 
\begin{equation}
Q_V=\frac{i}{2}\sum_x \lambda^1(x)\lambda^2(x)
\end{equation}
For infinitesimal $\theta$ the transformation becomes
\begin{equation}
    \lambda^a(x)\to \lambda^a-i\theta[Q_V,\lambda^a]
\end{equation}
From the fundamental anticommutators of the fields one deduces
\begin{align}
    \lambda^1(x)&\to \lambda^1(x)-\theta\lambda^2(x)\nonumber\\
    \lambda^2(x)&\to \lambda^2(x)+\theta\lambda^1(x)
\end{align}
It is easily verified that $[H,Q_V]=[H,U_V]=0$ as expected.

For the rest of
this section we will again focus on the interesting case of $d=3$ although it is
not hard to generalize the results to arbitrary dimension $d$.
We have already seen that a discrete $U_A(1)$ transformation corresponds to right multiplication of the matrix fermion $\Lambda$ by $\gamma_5$ and maps to the 3-shift $S_{123}$ operator 
acting on the staggered field. 
Its axial character can be made obvious by observing that
\begin{equation}
    \Lambda\gamma_5=(\gamma_5\Lambda\gamma_5)\gamma_5=\gamma_5\Lambda
\end{equation}
where we have used the fact that $\Lambda$ commutes with
$\gamma_5$ or equivalently has eigenvalue unity
under the twisted chiral operator $\gamma_5\otimes \gamma_5$. It is
natural to look for a generalization of the vector charge that incorporates this 3-shift, generates a continuous axial 
symmetry of the lattice system and becomes the singlet axial
charge in the continuum limit. In terms of the doublet $\lambda(x)$ 
it is given by
\begin{align}
   Q_A&=B_{123}Q_VB^{-1}_{123}\nonumber\\
   &=\frac{i}{2}\sum_x 
   \Xi_{123}(x)\lambda^1(x)\lambda^2(x+\hat{1}+\hat{2}+\hat{3})
\end{align}
where $\Xi_{123}(x)=\xi_1(x)\xi_2(x+\hat{1})\xi_3(x+\hat{1}+\hat{2})$ is the phase associated with the 3-shift. It is very important to note that here, following \cite{Chatterjee:2024gje}, we have employed a 3-shift only on the
$\lambda^2$ field i.e the shift is not a full shift $S_{123}$ but just $B_{123}$. A simultaneous shift on both fields would leave the vector charge invariant. Since $B_{123}$ is a symmetry of the (massless) Hamiltonian the resultant
charge $Q_A$ also commutes with the Hamiltonian.~\footnote{Start with $HQ_V=Q_VH$. Multiply left and right by $B_{123}$ and $B_{123}^{-1}$ respectively and use the fact that $B_{123}$ commutes with $H$ to show that $B_{123}Q_VB_{123}^{-1}$ must also commute with $H$.} Again, we can exponentiate this charge to produce a continuous
lattice symmetry corresponding to the operator $U_A=e^{-i\theta Q_A}$.
To see this note that 
\begin{align}
U_A&=e^{\theta/2\sum_x \Xi_{123}(x)\lambda^1(x)\lambda^2(x+\hat{1}+\hat{2}+\hat{3})}\nonumber\\
&=\prod_x\,e^{\theta/2\,\Xi_{123}(x)\lambda^1(x)\lambda^2(x+\hat{1}+\hat{2}+\hat{3})}\nonumber\\
=\prod_x&\left(\cos{\left(\frac{\theta}{2}\right)}
+\sin{\left(\frac{\theta}{2}\right)}\,\Xi_{123}(x)\lambda^1(x)\lambda^2(x+\hat{1}+\hat{2}+\hat{3})\right)
\end{align}
This acts as a rotation on the doublet ($U_A s^a U_A^\dagger$ where, $a \in \{1,2\}$)
    \begin{align}
    s^1(x)&=\lambda^1(x)\nonumber\\
    s^2(x)&=\Xi_{123}(x)\lambda^2(x+\hat{1}+\hat{2}+\hat{3})
    \end{align}
giving
\begin{align}
    s^1(x)&\to \cos{(\theta)}s^1(x)-\sin{(\theta)} s^2(x)\nonumber\\
    s^2(x)&\to \cos{(\theta)}s^2(x)+\sin{(\theta)} s^1(x)
\end{align}
However, if we
compute the commutator with the vector charge we find a non-zero result
\begin{align}
    [Q_V, Q_A]  & = -\frac{1}{2}\sum_{x}\Xi_{123}(x)\left[\lambda^1(x)\lambda^1(x+\hat{1}+\hat{2}+\hat{3})\right.\nonumber\\
    &\left. \quad -\lambda^2(x)\lambda^2(x+\hat{1}+\hat{2}+\hat{3})\right]
\end{align}
While this clearly vanishes in the naive continuum limit
it is clearly non-zero on a finite lattice and suggests that the theory suffers from a mixed 't Hooft anomaly - when the lattice vector symmetry is gauged, $Q_A$ is broken. 

By following this strategy it should be clear that one can construct
a series of conserved and quantized charges for the remaining shift symmetries $B_a$ and $B_{ab}$ 
\begin{equation}
   Q_{\ha}=B_{a}Q_VB_{a}^{-1} \quad Q_{\ha+\hb}=B_{ab}Q_V B_{ab}^{-1}\quad (d=3)
\end{equation}
These, again, can be exponentiated to yield continuous symmetries $U_a=e^{-i\theta Q_{\ha}}$ and $U_{ab}=e^{-i\theta Q_{\ha+\hb}}$.
However, these charges will not in general commute with the vector symmetry
\begin{equation}
    [Q_V,Q_{\ha}]=G_{\ha}\quad [Q_V,Q_{\ha+\hb}]=G_{\ha+\hb}
\end{equation}
where 
\begin{equation}
    G_{\vn}=-\frac{1}{2}\sum_x \Xi_{\vn}(x)\left(\lambda^1(x)\lambda^1(x+\vn)-\lambda^2(x)\lambda^2(x+\vn)\right)\nonumber\\
\end{equation}
where $\Xi_{\vn}(x)$ is the phase
associated with the shift $\vn$. Thus in three (spatial) dimensions $\Xi_{\vn}(x)=\xi_a(x)$ or $\xi_a(x)\xi_b(x+\vec{a})$ or $\xi_1(x)\xi_2(x+\hat{1})\xi(x+\hat{1}+\hat{2})$. $G_{\vn}$ commutes with $H$ and satisfies $G_{-\vn}=-G_{\vn}$.

In a similar fashion, one can compute the commutators of these charges with $Q_A$. For example using
\begin{align}
    Q_AQ_{\hat{1}}&=B_1B_2B_3Q_VB_3^{-1}B_2^{-1}B_1^{-1}B_1Q_VB_1^{-1}\nonumber\\
    &=B_1Q_{\hat{2}+\hat{3}}Q_VB_1^{-1}
\end{align} we find
\begin{equation}
    [Q_A,Q_{\hat{1}}]=-B_1G_{\hat{2}+\hat{3}}B_1^{-1}
\end{equation}
Similarly 
\begin{align}
  Q_AQ_{\hat{1}+\hat{2}}&=B_1B_2B_3 Q_VB_3^{-1}B_2^{-1}B_1^{-1}B_1B_2Q_VB_2^{-1}B_1^{-1}\nonumber\\
    =&B_1B_2 Q_{\hat{3}}Q_V B_1^{-1}B_2^{-1}
\end{align}
leading to
\begin{equation}
    [Q_A,Q_{\hat{1}+\hat{2}}]=-B_1B_2G_{\hat{3}}B_1^{-1}B_2^{-1}
\end{equation}

Other commutators follow a similar pattern eg
\begin{equation}
    [Q_{\hat{1}},Q_{\hat{3}}]=B_1G_{\hat{3}-\hat{1}}B_1^{-1}
\end{equation}
and
\begin{align}
    [Q_{\hat{2}+\hat{1}},Q_{\hat{2}+\hat{3}}]&=[B_2B_1Q_VB_1^{-1}B_2^{-1},B_2B_3Q_VB_3^{-1}B_2^{-1}]\nonumber\\
    &=B_2[Q_{\hat{1}},Q_{\hat{3}}]B_2^{-1}=B_2G_{\hat{3}-\hat{1}}B_2^{-1}
\end{align}
and
\begin{equation}
    [Q_{\hat{1}},Q_{\hat{1}+\hat{2}}]=B_1[Q_V,Q_{\hat{2}}]B_1^{-1}=B_1 G_{\hat{2}}B_1^{-1}
\end{equation}
and
\begin{equation}
    [Q_{\hat{1}},Q_{\hat{2}+\hat{3}}]=[Q_{\hat{1}},Q_{\hat{2}+\hat{3}-\hat{1}+\hat{1}}]=B_1G_{\hat{2}+\hat{3}-\hat{1}}B_1^{-1}
\end{equation}

These commutation relations can be summarized as
\begin{equation}
    [Q_{\vn},Q_{\vm}]=\hat{G}_{\vm-\vn}^{\vec{M}}
\end{equation}
where $\hat{G}_{\vn}^{\vM}=B_{\vM} G_{\vn} B_{\vM}^{-1}$ and $B_{\vM}$ denotes
a half shift corresponding to the vector $\vM$. The latter has non-zero components arising from repeated elementary vectors in $\vn$ and $\vm$. For example if $\vn=\hat{1}+\hat{2}$ and $\vm=\hat{1}+\hat{2}+\hat{3}$ then
$\vM=\hat{1}+\hat{2}$ and $B_{\vM}=B_{12}=B_1B_2$. In detail
\begin{align}
\hat{G}_{\vn}^M&=-\frac{1}{2}\sum_x \Xi_{\vn}(x)\left[\lambda^1(x)\lambda^1(x+\vn)\right.\nonumber\\
&\left.-\Xi_{\vec{M}}(\vn)\lambda^2(x+\vec{M})\lambda^2(x+\vn+\vec{M})\right]\end{align}
which simplifies to
\begin{align}
\hat{G}_{\vn}^M&=-\frac{1}{2}\sum_x \Xi_{\vn}(x)\left[\lambda^1(x)\lambda^1(x+\vn)\right.\nonumber\\
&\left.-\Xi_{\vn}(\vec{M})\Xi_{\vec{M}}(\vn)\lambda^2(x)\lambda^2(x+\vn)\right]\end{align}

There are additional non-trivial commutation relations between the $Q$ and $\hat{G}$ operators, and, if one allows for multiples of the elementary
shifts, the algebra described here
generalizes the Onsager algebra described in \cite{Chatterjee:2024gje} to higher dimensional lattices. We postpone a detailed
investigation of this  algebra to future work and merely
note here that the representation theory of this non-abelian algebra would provide a set of maximally commuting operators and thereby encode any continuum anomaly.

It is interesting to compute the commutator of $G_{\vn}$ with
the fields $\lambda^a$. One finds
\begin{align}
[G_{\vn},\lambda^1(x)]&=\Xi_{\vn}(x)\left(\lambda^1(x+\vn)-\lambda^1(x-\vn)\right)\nonumber\\
[G_{\vn},\lambda^2(x)]&=-\Xi_{\vn}(x)\left(\lambda^2(x+\vn)-\lambda^2(x-\vn)\right)
\end{align}

In momentum space the commutator behaves as
\begin{equation}
    [G_{\vn},\lambda^a(\vec{k})]\sim \left(
e^{i\vec{k}.\vn}-e^{-i\vec{k}.\vn}
    \right)\lambda^a(\vec{k})
    \label{kspace}
\end{equation} 
where $\vec{k}=\frac{2\pi m}{L}\quad{\rm with}\;m=-\frac{L}{2}\ldots \frac{L}{2}$ for
periodic boundary conditions.

This vanishes on zero energy modes of the
form 
\begin{equation}
    \lambda^a(x)\sim \frac{1}{\sqrt{V}}e^{i\pi\vec{A}^a.x}\label{doubler}
\end{equation}
where $\vec{A}$ is one of the shift vectors whose components take the values $0,1$ and $V$ is the lattice volume. The non-trivial $\vec{A}$ are the doublers.
On low energy modes in the vicinity of these zero energy states
the right hand of eqn~\ref{kspace} scales like $1/L$ and hence vanishes as $L\to\infty$. The fact that $G_{\vn}$ commutes with low energy modes of the field operator
at large $L$ and annihilates the vacuum
can be used to show that the matrix element of $G_{\vn}$ (and $\hat{G}_{\vn}^{\vM}$) between any two low energy states goes to zero in the naive continuum limit following the argument given in \cite{Chatterjee:2024gje}.

\section{Conclusions}
In this paper we have examined the shift, time reversal and phase
symmetries of
Hamiltonian staggered fermions on finite spatial lattices focusing
on the case of 3+1 dimensions. 
We have \SC{reviewed}
how the shift symmetries correspond to a discrete subgroup of the product
of the continuum axial-flavor symmetry and translations. In particular, the
odd shifts correspond to discrete axial transformations in the continuum theory
and are only symmetries in the massless theory. Furthermore, we find
that the odd shifts anticommute with time reversal.

We have constructed explicit operators to generate these symmetries along the lines of \cite{Seiberg:2023cdc}. To do this we decompose the complex staggered
fields into two real fields. This also enlarges the set of shift symmetries of the massless theory - one
can apply independent half shifts to each of these two fields. The presence of anticommuting symmetries hints
at the presence of 't Hooft anomalies. However, we have argued that
the system can be gapped, and hence the 
't Hooft anomaly canceled, for four complex staggered fields yielding sixteen Majorana fermions in three (spatial) dimensions. We conjecture that canceling these mixed shift-time reversal anomalies in the Hamiltonian 
formalism may be equivalent to canceling the gravitational anomalies of the Euclidean theory.

We have also constructed a set of local, conserved and quantized charges by combining each of the half shift symmetries with the vector charge $Q_V$. The resultant
charges generate a set of new continuous global symmetries of the lattice
theory. However, these charges do not
commute with the vector charge on finite lattices.  This implies that gauging the lattice vector symmetry will break these global symmetries - a lattice 't Hooft anomaly. In addition the commutation relations of the full set of charges form
a non-trivial algebra. We plan to investigate the representations of this non-abelian algebra in future work. This
will reveal the maximal set of commuting operators which should determine 
the continuum anomalies that can arise as the continuum
limit is taken.

While writing this paper we became aware of another recent work which also
elucidates
the symmetry structure of Hamiltonian staggered fermions with the
goal of classifying the possibilities for symmetric
mass generation \cite{Li:2024dpq}. Our results are consistent with their conclusions
where the two papers overlap.
 
\section*{Acknowledgements}
This work was supported by the US Department of Energy (DOE), 
Office of Science, Office of High Energy Physics, 
under Award Number DE-SC0009998. This research 
was supported in part by grant NSF PHY-2309135 to the Kavli Institute for Theoretical Physics (KITP). SMC would like to acknowledge useful conversations with Theo Jacobson and Thomas Dumitrescu. 

\begin{appendix}
\begin{comment}
\section{Energy level degeneracy}
Consider a system which is invariant under time reversal $\T$ and 
another symmetry $A$ with $A$ and $\T$ anticommuting. Thus
\begin{equation}
    [H,A]=0\quad [H,\T]=0 \quad {\rm with} \quad \{A,\T\}=0
\end{equation}
The eigenvectors of $H$ can be labeled with say the eigenvalues $\lambda_A$ of $A$ 
\begin{equation}
    H\ket{\lambda_A}=E\ket{\lambda_A}
\end{equation}
Acting on the left with $\T$ we find that 
\begin{equation}
    H\,\T\ket{\lambda_A}=E\,\T\ket{\lambda_A}
\end{equation}
Thus the state $\T\ket{\lambda_A}$ also has energy $E$. To show that it is not proportional to $\ket{\lambda_A}$ consider
\begin{equation}
    \bra{\lambda_A}(A\T+\T A)\ket{\lambda_A}=0
\end{equation}
But this implies
\begin{equation}
    (\lambda_A+\lambda_A^*)\braket{\lambda_A \vert \T\lambda_A}=0
\end{equation}
\SC{Hence $\ket{\lambda_A}$ and $\T\ket{\lambda_A}$ are orthogonal if ${\rm Re}\,\lambda_A\ne 0$. Since $\lambda_A$ is a pure phase this proof will
fail if $\lambda_A=\pm i$. In our case $A$ corresponds to
a shift operator $S_i$ whose phase can be chosen such that $S_i^2=T_i$. Assuming the ground state is invariant under translations we infer that
$T_i\ket{0}=\ket{0}$ which implies that $\lambda_A=\pm 1$ and hence the ground state will be doubly degenerate}
\end{comment}
\SC{
\section{Connections between Euclidean and Hamiltonian staggered fermions}
Since a single Dirac fermion has $2^{\frac{d+1}{2}}$ components, the most general square matrix that can be built from such spinors will possess
$2^{\frac{d+1}{2}}\times 2^{\frac{d+1}{2}}=2\times 2^d$ elements.
This is twice the number of points in the spatial cube 
employed in our Hamiltonian construction and hence cannot be reproduced 
with our single Hamiltonian staggered fermion. Clearly the missing degrees of
freedom correspond to the fact that we truncated the theory before spin diagonalization by retaining only fields $\chi_+(x)$ whose $\beta$-eigenvalue
was equal to unity. Equivalently the matrix fermion we construct commutes
with $\gamma_5$. It corresponds to a reduced staggered fermion in the
Euclidean formulation \cite{vandenDoel:1983mf,Catterall:2015zua,Catterall:2023nww}. 
Indeed, the mass term that arises in the Hamiltonian formulation can be
seen as the dimensional reduction of the temporal one-link mass term
of a reduced fermion.
}

\SC{
To achieve the full set of Dirac
fermions 
one can employ a second staggered fermion $\chi^\prime$ and expand it on the additional matrix basis given by $\beta\alpha^x$
\begin{equation}
\Lambda^\prime(x)=\frac{1}{8}\sum_{\{b\}}\chi^\prime(x+b)\beta\alpha^{x+b}
\label{otherpsi}
\end{equation}
This matrix fermion anticommutes with $\gamma_5$.
By adding $\Lambda$ and $\Lambda^\prime$ we can then build a theory
containing $2^{\frac{d+1}{2}}$ Dirac fermions corresponding to
the number of fermions that arise in Euclidean
formulations of full staggered fermions where time is also
discretized.}

\section{Additional properties of staggered fermions}
In this appendix we list the additional symmetries of the free staggered fermion Hamiltonian that are not already covered in the main text \cite{Golterman:1984cy,Golterman:2024xos}.
\subsection{Rotational Invariance}
Consider the transformation
\begin{align*}
\chi(x) \rightarrow S_R(R^{-1}x)\chi (R^{-1}x)
\end{align*}
where the rotation matrix \( R \equiv R^{pr}\) acts on the spatial coordinates as
\[
x_p \to x_r, \quad x_r \to -x_p, \quad x_s \to x_s \quad (s \neq p, r)
\]
and
\begin{align}
    S_R(x) = &\frac{1}{2} [ 1 \pm \eta_p(x) \eta_r(x) \mp \xi_p(x) \xi_r(x) \\
    &\quad+ \eta_p(x) \eta_r(x) \xi_p(x) \xi_r(x)] \ \ , \ p\lessgtr r
\end{align}
Invariance of the Hamiltonian follows from 
\begin{equation}
   S_R(R^{-1}x) \eta_i(x) S_R(R^{-1}x + R^{-1}(x+i)) = R_{ij} \, \eta_j (R^{-1}x) 
\end{equation}
\subsection{Axis reversal}
We define an axis reversal transformation \( I \equiv I^{(p)} \), which acts on the spatial coordinates as
\begin{equation}
    x'_i = 
\begin{cases}
- x_i & \text{if } i = p \\
\phantom{-} x_i & \text{if } i \ne p
\end{cases}
\end{equation}
Under this transformation, the field transforms as
\begin{align*}
\chi(x) \rightarrow (-1)^{x_p} \chi(Ix)  
\end{align*}

\subsection{Spin-Taste basis}
\SC{To understand the continuum limit of staggered fermions it is
useful to employ what has been termed ``taste basis" for
Euclidean staggered fermions \cite{Golterman:2024xos}. For the Hamiltonian
theory we can provide an analogous construction by defining two matrix
fermions on a lattice with twice the lattice spacing:
\begin{align}
\psi^+_{ba}(y) &=\frac{1}{2^{1/2}}\sum_{A\in A_e} (\alpha^A)_{ba} \, \chi_e(2y + A)\nonumber\\
\psi^-_{ba}(y) &=\frac{1}{2^{1/2}}\sum_{A\in A_o} (\alpha^A)_{ba}\, \chi_o(2y+A)\label{spintaste}
\end{align}
where $A$ denotes a unit cell vector with components \( 0 \) or \( 1 \) and
$A_e$ denotes the subset of such vectors with an even
number of non-zero components while $A_o$ denotes those with an odd number. As before
\begin{equation}
\alpha^A = \alpha_1^{A_1} \alpha_2^{A_2}\alpha_3^{A_3} 
\end{equation}
The matrix index \( b \) is interpreted as a ``spin" index while \( a \) is interpreted as a ``taste" index. Notice that both $\psi_+$ and $\psi_-$
are constructed from just four complex parameters and hence the field $\psi=\psi_++\psi_-$ can contain only two Dirac fermions as noted in the main text. Another way to see this is to note that $\psi$ commutes with $\gamma_5=i\alpha_1\alpha_2\alpha_3$.
Since $\beta$ commutes with $\alpha^{A_e}$ and anticommutes with $\alpha^{A_o}$ the matrix fields on the left satisfy 
\begin{align}
    \beta\psi_+\beta&=\psi_+\nonumber\\
    \beta\psi_-\beta&=-\psi_-
\end{align}
Eqn.~\ref{spintaste} can be inverted (using $\text{tr}(\alpha^A\alpha^B)=4\delta^{AB}$) to give
\begin{align}  
\chi(2y+A)=\frac{1}{2^{3/2}}\text{tr}\left(\frac{1}{2}(\psi(y)+\epsilon(A)\beta\psi(y)\beta)\alpha^A\right)
\end{align}
The mass term becomes
\begin{align}
    \sum_x\epsilon(x)\chi^\dagger(x)\chi(x)&=\frac{1}{8}\sum_{y,A}\text{tr}(\psi^\dagger(y)\alpha^A)\text{tr}(\beta\psi(y)\beta\alpha^A)\nonumber\\
    &=\frac{1}{2}\sum_y \text{tr}(\psi^\dagger\beta\psi(y)\beta)
\end{align}
where we have used the completeness relation 
\begin{equation}
    \sum_A \alpha^A_{ab}\alpha^A_{cd}=4\delta_{ad}\delta_{bc}
\end{equation}
which is valid for any matrix which can be
expanded on the $\alpha^A$.
Using the results
\begin{align}
    \sum_{A,A_i=0}\eta_i(A)(\alpha^{A+i})_{\alpha a}(\alpha^A)_{b\beta}&=4(\alpha_i)_{\alpha\beta}\delta_{ab}+4\beta_{\alpha\beta}(\beta\alpha_i)_{ba}\nonumber\\
    \sum_{A,A_i=0}\eta_i(A)(\alpha^{A})_{\alpha a}(\alpha^{A+i})_{b\beta}&=4(\alpha_i)_{\alpha\beta}\delta_{ab}-4\beta_{\alpha\beta}(\beta\alpha_i)_{ba}
\end{align}
one can show that the Hamiltonian in the spin-taste
basis is given by
\begin{align}
    H &=\frac{i}{4}\sum_{y, i} \text{tr}\left[ 
\psi^\dagger(y)\alpha_i\left(\psi(y + i) - \psi(y - i) \right)\right]\nonumber\\
&+\text{tr}\left[\psi^\dagger(y)\beta\left( \psi(y + i) 
+ \psi(y-i)-2\psi(y)\right)\beta \alpha^i\right]\nonumber\\ 
&+\frac{1}{2}\sum_{y}\text{tr}(\psi^\dagger(y)\beta\psi(y)\beta)
\end{align}
%\textcolor{teal}{will recheck}
We see that in this form, the free Hamiltonian describes two naive Dirac fermions, with the addition of a non-diagonal Wilson-like mass term, which removes the doublers. 
}

\end{appendix}
\bibliography{refs}

\end{document}